\documentclass[prb,superscriptaddress,noshowpacs,twocolumn,reprint,noshowkeys,longbibliography,amssymb,amsmath,amsfonts]{revtex4-1}

\usepackage{graphicx}
\usepackage{theorem}
\usepackage{dcolumn}
\usepackage{longtable}
\usepackage{bm}
\usepackage{color} % textcolor

\newcommand{\ie}{{\textit{i.e.}}}
\newcommand{\eg}{{\textit{e.g.}}}
\newcommand{\cf}{{\textit{cf.}}}

\graphicspath{{Fig/}}

\newcommand{\tc}{T_{\text{c}}}
\newcommand{\jc}{j_{\text{c}}}
\newcommand{\fmod}{f_{\text{mod}}}
\newcommand{\Tmod}{T_{\text{mod}}}
\newcommand{\Amod}{A_{\text{mod}}}
\newcommand{\fres}{f_{\text{res}}}
\newcommand{\fpl}{f_{\text{p}}}

\newcommand{\lj}{\lambda_{\mathrm{J}}}

\newcommand{\co}[2]{\ifcase #1 \or #2 \fi}

\newcommand{\TB}{T_{\text{b}}}
\renewcommand{~}{\,}

\makeatletter
\renewcommand{\figurename}{Fig.}
\renewcommand{\tablename}{Tab.}
\renewcommand{\fnum@figure}[1]{\textbf{\figurename~\thefigure} }
\renewcommand{\fnum@table}[1]{\textbf{\tablename~\thetable} }
\makeatother

\begin{document}

\title{Space-time crystalline order of a high-critical-temperature superconductor with intrinsic Josephson junctions}

\author{Reinhold Kleiner}
\email{kleiner@uni-tuebingen.de}
\affiliation{Physikalisches Institut, Center for Quantum Science (CQ) and LISA$^+$, Universit\"{a}t T\"{u}bingen, D-72076 T\"{u}bingen, Germany}

\author{Xianjing Zhou}
\affiliation{Center for Nanoscale Materials, Argonne National Laboratory, Argonne, Illinois 60439, USA}

\author{Eric Dorsch}
\affiliation{Physikalisches Institut, Center for Quantum Science (CQ) and LISA$^+$, Universit\"{a}t T\"{u}bingen, D-72076 T\"{u}bingen, Germany}

\author{Xufeng Zhang}
\affiliation{Center for Nanoscale Materials, Argonne National Laboratory, Argonne, Illinois 60439, USA}

\author{Dieter Koelle}
\affiliation{Physikalisches Institut, Center for Quantum Science (CQ) and LISA$^+$, Universit\"{a}t T\"{u}bingen, D-72076 T\"{u}bingen, Germany}

\author{Dafei Jin}
\email{djin@anl.gov}
\affiliation{Center for Nanoscale Materials, Argonne National Laboratory, Argonne, Illinois 60439, USA}

\date{\today}% It is always \today, today,

\begin{abstract}
We theoretically demonstrate that the high-critical-temperature superconductor Bi$_2$Sr$_2$CaCu$_2$O$_{8+x}$ (BSCCO) is a natural candidate for the recently envisioned classical space-time crystal. BSCCO intrinsically forms a stack of Josephson junctions. Under a periodic parametric modulation of the Josephson critical current density, the Josephson currents develop coupled space-time crystalline order, breaking the continuous translational symmetry in both space and time. The modulation frequency and amplitude span a (nonequilibrium) phase diagram for a so-defined spatiotemporal order parameter, which displays rigid pattern formation within a particular region of the phase diagram. Based on our calculations using representative material properties, we propose a laser-modulation experiment to realize the predicted space-time crystalline behavior. Our findings bring new insight into the nature of space-time crystals and, more generally, into nonequilibrium driven condensed matter systems.

\end{abstract}

\pacs{74.50.+r, 74.72.-h, 85.25.Cp}

\maketitle

\pretolerance=8000 % good to put here,otherwise may change the affiliation style

In recent years, the notion of a space-time crystal (STC) has attracted a great deal of attention \cite{Wilczek12,Shapere12,Li12,Bruno13,Watanabe15b,Lazarides14,sacha2015modeling,Lazarides15,Khemani16,yao2017discrete,Xu18,Sacha18,kozin2019quantum}. While there has been considerable discussion about what is truly outstanding in such a system, at present, it is mostly agreed that the STC refers to a nonequilibrium phase of matter displaying long-range order in both space and time \cite{Xu18}. Specifically, the (nonlinear) many-body interaction makes the system exhibit long-lived oscillations at a period longer than the period of the driving source, and the oscillation patterns show rigidity against perturbation from the environment \cite{sacha2015modeling,else2016floquet,yao2017discrete,Sacha18}. These extraordinary behaviors have been theoretically conceived and experimentally observed in atomic-molecular-optical (AMO) systems \cite{Smits18,Zhang17,Li12,Choi17,Rovny18,pal2018temporal}. But few candidates from condensed matter systems have been explored \cite{autti2018observation,Homann2020PRR}.

\begin{figure}[htb]
\centerline{\includegraphics[scale=0.65]{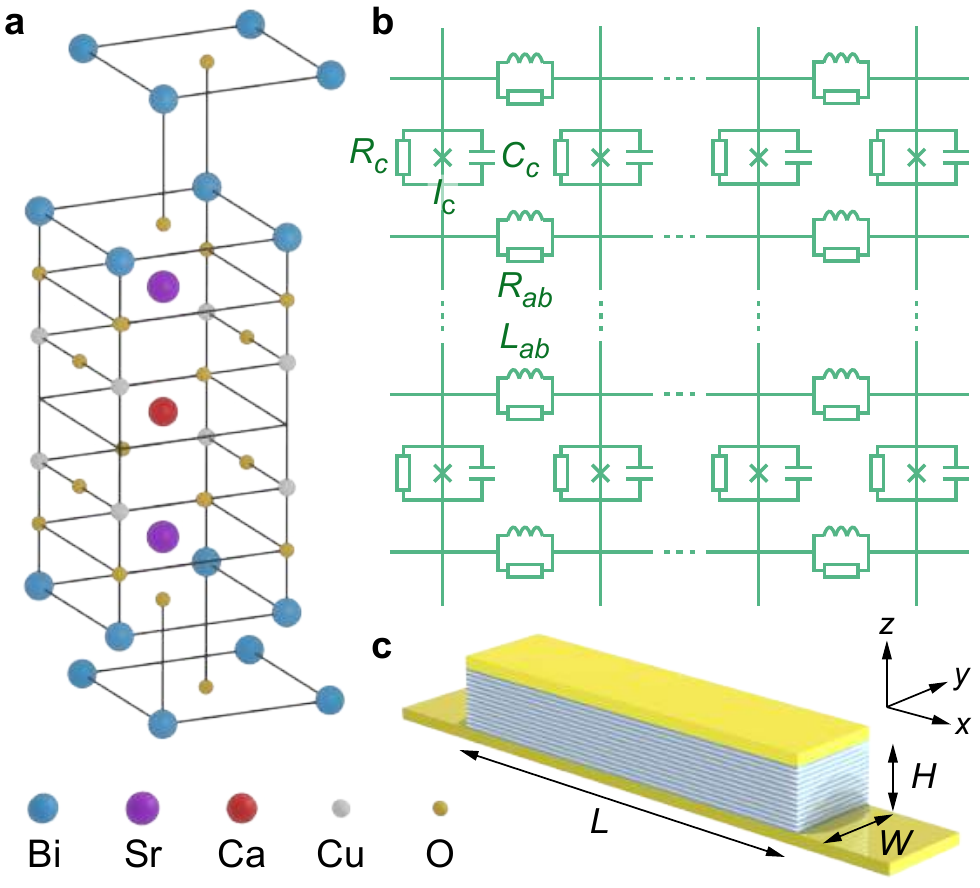}}
\caption{\textbf{Schematic of the superconductor BSCCO with intrinsic Josephson junctions.} \textbf{a} Atomic structure in which the CuO$_2$ planes are superconducting. \textbf{b} Equivalent circuit. Each discretized element is characterized by the Josephson critical current $I_\text{c}$, $c$-axis resistance $R_c$, $c$-axis capacitance $C_c$, $ab$-plane resistance $R_{ab}$, and $ab$-plane inductance $L_{ab}$ which is associated with the in-plane supercurrents and dominantly kinetic in origin. Additional geometric inductances could be added in series to $R_{ab}$ but are neglected. The discreteness along $z$ is naturally set by the layered structure, whereas the element sizes along $x$ and $y$ are in theory infinitesimal, $dx, dy\rightarrow 0$. In numerical simulation, $dx$ and $dy$ are finite but smaller than the Josephson length $\lj$. Here we chose $dx = dy = 0.5~\mu$m, which is roughly a factor of 2 below $\lj$. \textbf{c} A typical rectangular-shaped 3D junction stack of length $L$, width $W$, and height $H$.}
\label{fig:schematic}
\end{figure}

In this paper, we present theoretically that the high critical temperature (high-$T_\text{c}$) cuprate superconductor Bi$_2$Sr$_2$CaCu$_2$O$_{8+x}$ (BSCCO) is a natural candidate of a classical discrete STC that was recently envisioned \cite{Yao20}. This material, as illustrated in Figure~\ref{fig:schematic}, acts as a stack of intrinsic Josephson junctions (IJJs) along its crystallographic $c$ axis with $s=1.5$~nm layer period  \cite{Kleiner92,Hu10,Savelev10}. Each junction is formed by the insulating BiO and SrO planes sandwiched between the superconducting CuO$_2$ planes. In the crystallographic $ab$ plane, the junctions can be macroscopically large ($\gtrsim1$~mm) and behave as a long (and wide) Josephson junction. Adjacent junctions are coupled via phase gradients of the superconducting wave function, produced by currents flowing along the CuO$_2$ planes. The number $N$ of junctions in a stack can vary from a few to thousands.

Our calculations indicate that when the critical current density $\jc$ of the junctions is subject to a parametric modulation, which is made periodic in time and uniform in space, BSCCO spontaneously develops half-harmonic oscillations of the Josephson currents in time and broken (continuous) translational symmetry in space. The newly formed order goes beyond a direct product of separate spatial and temporal orders and embodies space-time coupled symmetry \cite{Xu18}. At a fixed temperature $T$ and thermal noise, the modulation frequency and amplitude span a phase diagram. Within a specific region with clear boundaries in this phase diagram, a nonzero spatiotemporal order parameter emerges, indicating the necessary robustness in phase formation.

Although parametric driving and half-harmonic generation of nonlinear oscillators such as Josephson junctions or junction arrays have been investigated previously \cite{Feldman75, Pedersen80, Pedersen00}, their conceptual connection to the space-time crystalline order was only recently proposed \cite{Yao20}. Our study verifies that half-harmonic generation is accompanied with the spatiotemporal order formation in any spatial dimensions of BSCCO. But most notably, the STC phase is stable only at nonzero spatial dimension; the phase rigidity appears to increase with the spatial dimensionality. Therefore, our study of the intrinsic Josephson junction stack in BSCCO suggests the possible realization of a classical discrete space-time crystal from a naturally existing condensed matter system.

\noindent\textbf{Results}

\noindent\textbf{Josephson plasma and photon modes in BSCCO.} The dynamics of a BSCCO stack follows the inductively coupled sine-Gordon equations \cite{Sakai93,Bulaevskii94,Rudau15, Rudau16}. These are nonlinear differential equations for the gauge-invariant Josephson phase differences, parametrically dependent on the Josephson critical current density $j_{\rm c}(T)$, the in-plane and out-of-plane resistivities $\rho_{ab}(T)$ and $\rho_{c}(T)$, and the Cooper pair density $n_{\rm s} (T)$. All of these quantities are $T$-dependent. The BSCCO stack supports collective oscillations of the supercurrents -- the Josephson-plasma oscillations -- across the junctions \cite{Tsui94,Matsuda95}. These oscillations can couple with the electromagnetic waves in the stack in the frequency range from $\lesssim0.1$~THz to $\gtrsim1$~THz \cite{Ozyuzer07,Savelev10,Hu10,Welp13,Kakeya16}. In practice, BSCCO stacks are routinely fabricated into rectangular bars with tens-of-micron length $L$ along $x$, several-micron width $W$ along $y$, and a height $H=Ns$ associated with tens or hundreds of junctions along $z$ (\cf\ Fig.~\ref{fig:schematic}). This stack, resembling a slab waveguide cavity, hosts a series of resonance modes at gigahertz-to-terahertz frequencies. For not too large oscillating amplitudes, the electric field and tunneling current along $z$ on the $n$th junction take the form \cite{Kleiner94,Sakai94},
\begin{equation}
E_z, j_z \propto \cos\left(\frac{\pi lx}{L}\right) \cos\left(\frac{\pi my}{W}\right) \sin\left(\frac{\pi q n}{N+1}\right),\label{eq:plasma_field}
\end{equation}
where $x\in[0,L]$, $y\in[0,W]$, and $n = 1, 2, \dots, N$. The mode indices $l$ and $m$ take zero or any positive integers, and $q$ takes a positive integer from $1$ to $N$. The associated resonance frequency reads
\begin{equation}
\label{eq:plasma_reso}
\fres = \sqrt{\left(C\fpl\right)^2+c_q^2\left[\left(\frac{\pi l}{L}\right)^2+\left(\frac{\pi m}{W}\right)^2\right]},
\end{equation}
with $\fpl\propto \sqrt{j_\text{c}{(T)}}$ being the Josephson plasma frequency and $c_q = \bar{c}/\sqrt{1-2\bar{s}\cos(\pi q/(N+1))}$ being the effective  speed of light,  where $\bar{c}$ is the Swihart velocity and $\bar{s}\approx 0.5$ is the interlayer coupling constant.
$C \leq 1$ is a correction factor which we introduce here to account for a frequency shift generated by parametric driving.
The values of $\jc$, $\fpl$ and $\bar{c}$ depend strongly on the charge carrier concentration. For convenience we use
$\jc \approx 250$~A~cm$^{-2}$, $\fpl\approx 47$~GHz and $\bar{c}\approx3.3\times10^5$~m/s at 4~K, which are typical for slightly underdoped BSCCO \cite{Rudau15,Rudau16}. These quantities drop to zero at the critical temperature $\tc \approx 85$~K.

If $l=m=0$, $E_z$ and $j_z$ are uniform in the $x$ and $y$ directions, independent of the lateral size $L$ and $W$, and irrespective of the values of $n$ and $q$ in the thickness direction. When $L$ and $W$ are both small ($\lesssim2$~$\mu$m), the only modes in the accessible frequency range are the uniform modes. In this case, if $N=1$, we have an effectively 0-dimensional (0D) single-layer short junction. On the other hand, when $L$ is large ($\gtrsim10$~$\mu$m) but $W$ is small, we can differentiate two cases: If $N=1$ and $q=1$, we have an effectively 1D single-layer long junction; and if $N>1$ and $q=1,2,\dots,N$, we have an effectively 2D multi-layer long junction stack. In these situations, $l$ can take nonzero integers but $m$ remains zero. Lastly, when $L$ and $W$ are both large and $N>1$, we return to a 3D multi-layer long and wide junction stack. Below, we show the emergence of space-time crystalline order in BSCCO with increasing dimensionality from 0D to 3D, under a temporally periodic but spatially uniform parametric modulation to $\jc$. While the case of a 0D single-layer short junction is merely a replica of the children's swing problem \cite{Belyakov09,glendinning2020adaptive}, exhibiting half-harmonic generation in time, the involvement of spatial dimensions from 1D to 3D not only exhibits half-harmonic generation but also stabilizes the temporal pattern on a much shorter time scale, and moreover, results in spontaneous (continuous) translational symmetry breaking in space. The parametric driving bandwidth and the robustness to thermal noise enhance significantly with the increase of spatial dimension.

It is worthwhile to point out that there are three essential ingredients leading to the nontrivial behavior of this system: the discrete electromagnetic modes as described by Eq.~(\ref{eq:plasma_field}), the nonlinear Josephson coupling of the superconducting planes, and the parametric drive through a periodic modulation of the Josephson critical current density. Superconductivity plays a key role in all these ingredients. Referring to Fig.~\ref{fig:schematic}b, in the normal state when superconductivity disappears, the inductive elements $L_{ab}$ would be absent ($L_{ab}\rightarrow\infty$) and the Josephson critical currents would be zero. Then the equivalent circuit would turn into a highly damped resistor network with linear capacitive couplings between the metallic layers. The mode spectrum as in Eq.~(\ref{eq:plasma_reso}) would be highly damped and, formally, $\fres$ would go to zero. For purely linear oscillators, parametric excitation still works. However, the oscillation amplitude may grow unbounded. We have explicitly tested this for the parameters of our system in the 0D and 1D cases by linearizing the Josephson current-phase relation. We found that the Josephson plasma oscillations or the cavity resonances were either not excited at all or grew exponentially. The sinusoidal Josephson current-phase relation removes this divergence.

\begin{figure*}[htb]
\centerline{\includegraphics[scale=0.48]{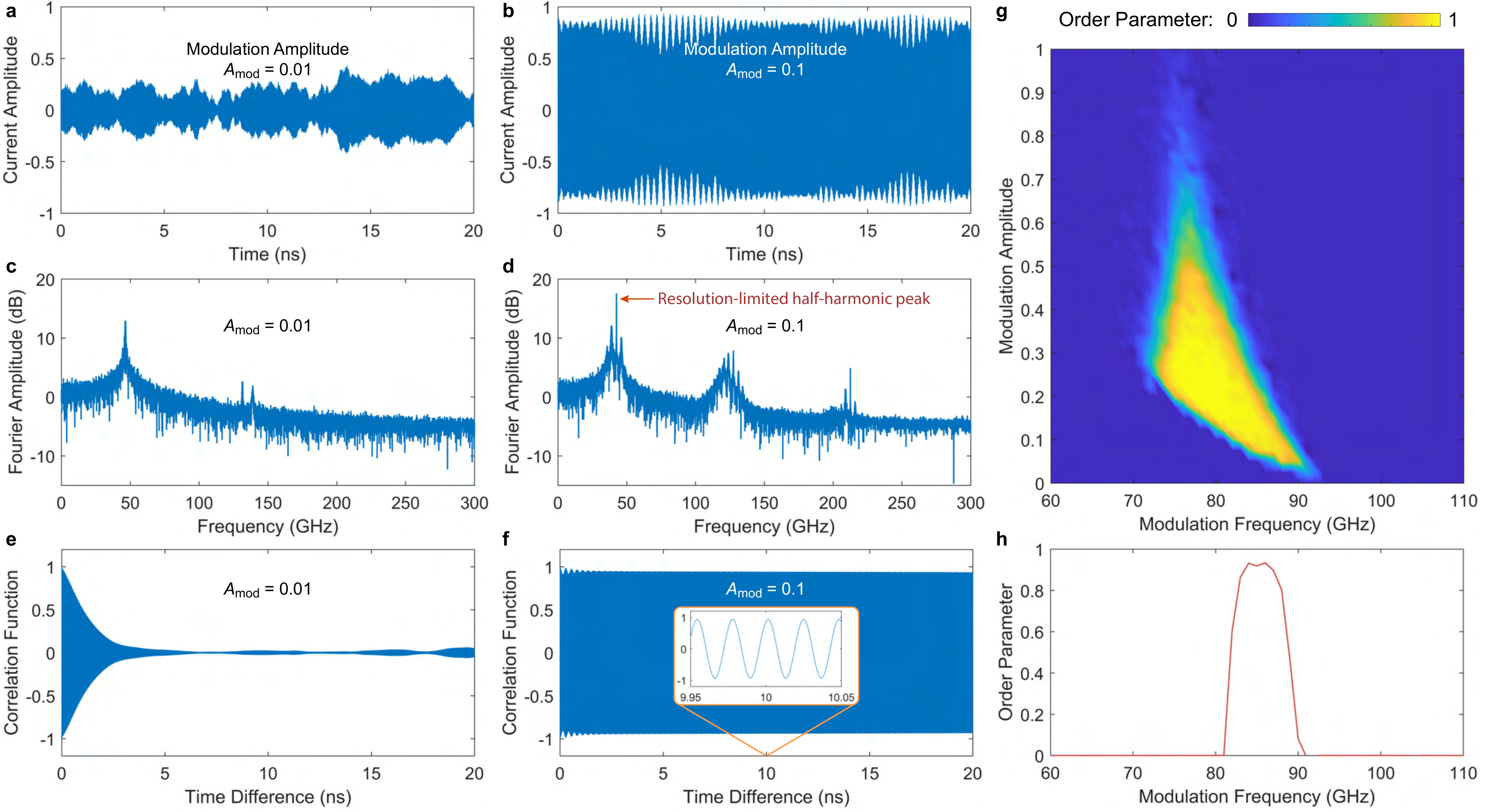}}
\caption{\textbf{Calculation on an effectively 0D single-layer short Josephson junction in BSCCO at 4~K under parametric modulation.} The junction area is $L\times W = 4$~$\mu$m$^2$ and the Josephson plasma frequency is $\fpl = 47$~GHz. \textbf{a} Time trace of the current density $j_z(t)$ under a modulation frequency $\fmod=85$~GHz and small modulation amplitude $\Amod=0.01$. The system does not converge to a Floquet steady state. \textbf{b} Time trace of $j_z(t)$ under a large modulation amplitude $\Amod=0.1$. The system has reached the Floquet steady state. \textbf{c} Fourier transform to $j_z(t)$ in \textbf{a}, plotted in the magnitude of $|j_z(f)|$. \textbf{d} Fourier transform to $j_z(t)$ in \textbf{b}, showing a resolution-limited sharp half-harmonic generation around the Josephson plasma frequency. \textbf{e} Time-averaged current-current correlation function $g(\delta t)$ for $\fmod=85$~GHz and small modulation amplitude $\Amod = 0.01$. \textbf{f} Time-averaged correlation function $g(\delta t)$ for a large modulation amplitude $\Amod=0.1$, showing an ordered temporal pattern. \textbf{g} Phase diagram of temporal order parameter $\Delta_t$ for an ensemble of 50 junctions versus the modulation frequency $\fmod$ and amplitude $\Amod$ (with fixed $\fpl$, temperature $T=4$~K, and thermal noise). Clear phase boundaries can be identified. \textbf{h} Temporal order parameter versus $\fmod$ when $\Amod=0.1$, showing a parametric excitation band within 80~--~90~GHz.}
\label{fig:0D}
\end{figure*}

\begin{figure*}[htb]
\centerline{\includegraphics[scale=0.5]{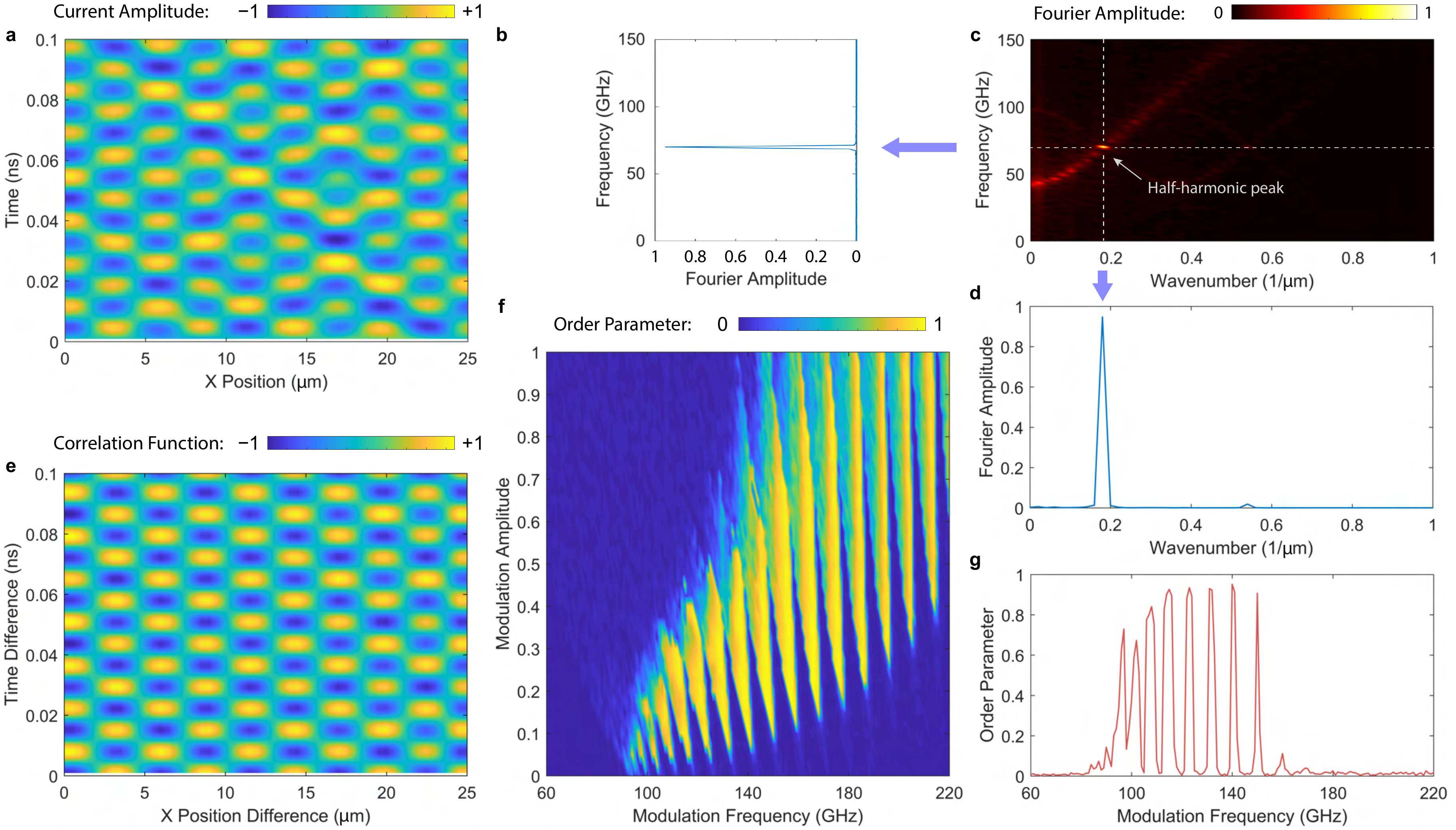}}
\caption{\textbf{Calculation on an effectively 1D single-layer long Josephson junction in BSCCO under parametric modulation at 4~K.} The junction length is $L=25$~$\mu$m and width is $W=2$~$\mu$m. \textbf{a} Space-time trace of the current density $j_z(t)$ in the Floquet steady state under the modulation frequency $\fmod=140$~GHz and modulation amplitude $\Amod=0.1$. \textbf{b} Linescan along the frequency axis in \textbf{c} at the peak wavenumber, showing a peak frequency around the half-harmonic frequency $\fmod/2\approx 70$~GHz. \textbf{c} Fourier transform to $j_z(x,t)$ in \textbf{a}, plotted in the magnitude $|j_z(\beta,f)|$, where $\beta$ is the wavenumber along $x$. The graph shows a strong half-harmonic peak in the wavenumber-frequency plane. \textbf{d} Linescan along the frequency axis in \textbf{c} at the peak frequency, showing a peak wavenumber corresponding to the mode index $l=9$. \textbf{e} Space-time-averaged current-current correlation function $g(\delta x,\delta t)$ for $\fmod=140$~GHz and $\Amod = 0.1$, showing an ordered spatiotemporal pattern. \textbf{f} Phase diagram of the spatiotemporal order parameter $\Delta_{xt}$ versus modulation frequency $\fmod$ and amplitude $\Amod$ (with fixed $\fpl$, temperature $T=$~4K, and thermal noise). Clear phase boundaries can be identified. \textbf{g} Spatiotemporal order parameter $\Delta_{xt}$ versus $\fmod$ when $\Amod=0.1$, showing a broadened parametric excitation band (compared with the 0D case) spanning 100 to 160~GHz, a range much higher than $2\fpl=94$~GHz. For $\Amod = 0.1$ the correction factor $C=0.95$.}
\label{fig:1D}
\end{figure*}

\noindent\textbf{Effectively 0D single-layer short junction.} Let us first look at an effectively 0D single-layer short junction, whose governing equation is identical to that of a nonlinear pendulum \cite{Stewart68, McCumber68}. If the critical current density $\jc$ is modulated in time, then the system is analogous to a children's swing. Our calculation takes account of the thermal (Nyquist) noise, which is related to the temperature $T$ by the fluctuation-dissipation theorem. In the presence of some initial fluctuations, the current density $j_z$ across the junction starts to oscillate, provided that the modulation frequency is about twice the intrinsic Josephson plasma frequency $\fpl$ and the modulation amplitude is large enough \cite{Belyakov09}. No inhomogeneous external force is applied to the system. Fig.~\ref{fig:0D} shows our calculation for a single-layer short junction of BSCCO with the junction area $L\times W = 4$~$\mu$m$^2$. The parametric modulation takes the form
\begin{equation}
\jc (T,t) = \jc(T) \frac{ 1+\Amod \cos(2\pi \fmod t) }{1+\Amod},\label{eqn:jc_mod}
\end{equation}
where $\jc(T)$ is the unmodulated critical current density, $\fmod$ is the modulation frequency and $\Amod$ is the modulation amplitude. Fig.~\ref{fig:0D}a and \ref{fig:0D}b show the time traces of $j_z(t)$, normalized to $\jc$ at 4~K, after many steps of initial relaxation with the same modulation frequency $\fmod=85~\text{GHz}\approx2\fpl$ but two different modulation amplitudes $\Amod=0.01$ (small) and 0.1 (large). For small modulation, the current amplitude is irregular. But for large modulation, the system can reach a Floquet steady state \cite{else2016floquet,oka2019floquet}, where the oscillation amplitude in time turns into an ordered pattern. Fig.~\ref{fig:0D}c and \ref{fig:0D}d give the Fourier transform of $j_z(t)$ in terms of the amplitude $|j_z(f)|$ for the two modulation cases. A sharp half-harmonic peak around $\fpl$ (along with a few other peaks at higher harmonics) can be observed in Fig.~\ref{fig:0D}d for large modulation.

To quantify the temporal order, we define a (dimensionless) current-current correlation function,
\begin{equation}
g(\delta t) = \frac{\langle \overline{j_z(t) j_z(t+\delta t)} \rangle_{t}}{\sqrt{\langle \overline{ j_z^2(t) } \rangle_{t} \langle \overline{j_z^2(t+\delta t)}\rangle_{t}}},
\end{equation}
where the time average is taken in the Floquet steady state. The overline denotes additional ensemble averaging. Fig.~\ref{fig:0D}e and \ref{fig:0D}f show $g(\delta t)$, which is irregular for small modulation amplitude and almost periodic (like a sinusoidal function, as shown in the inset of Fig.~\ref{fig:0D}f) for large modulation amplitude. To see the dependence of temporal order on the modulation parameters, we define a temporal order parameter,
\begin{equation}
\begin{split}
%\Delta_t = \frac{1}{2}\left\langle g(2\nu \Tmod)-g((2\nu+1) \Tmod) \right\rangle_\nu, \label{eq:OrderT}
& \Delta_{t} = \frac{1}{2} \left[ \frac{\langle j_z(t) j_z(t+2\nu \Tmod)\rangle_{t,\nu}}{\sqrt{\langle j_z^2 (t)\rangle_{t,\nu}\langle j_z^2 (t+2\nu \Tmod)\rangle_{t,\nu}}} \right.\\
& \left. - \frac{\langle j_z(t) j_z(t+(2\nu+1) \Tmod)\rangle_{t,\nu}}{\sqrt{\langle j_z^2 (t)\rangle_{t,\nu}\langle j_z^2 (t+(2\nu+1) \Tmod)\rangle_{t,\nu}}} \right] , \label{eq:OrderT}
\end{split}
\end{equation}
with the integer $\nu$ denoting even ($2\nu$) and odd ($2\nu+1$) number of modulation period $\Tmod$ in the time difference $\delta t$. The averaging is taken over both $t$ and $\nu$. The so-defined $\Delta_t$ emphasizes oscillations of $j_z$ occurring at twice the modulation time $2\Tmod=2/\fmod$, since the odd and even parts have the same moduli but different signs. By contrast, it suppresses oscillations that are either periodic in  $\Tmod$ or incommensurate. Fig.~\ref{fig:0D}g gives a ``phase diagram" of $\Delta_t$ versus $\fmod$ and $\Amod$ at the temperature $T=4$~K. We see that, like for the children's swing \cite{Belyakov09}, only if the modulation frequency falls in a narrow band near $2\fpl$ and the amplitude is neither too small or too large, the system can produce ordered half-harmonics. If the modulation amplitude is overly large, the system can become chaotic. Fig.~\ref{fig:0D}h is a linescan of Fig.~\ref{fig:0D}g at the modulation amplitude $\Amod=0.1$, showing a parametric excitation band within 80~--~90~GHz, near and slightly below $2\fpl=94$~GHz \cite{Belyakov09}.

An important observation unique to the 0D case is that the correlation function and order parameter appear to persistently decay even after exceedingly longtime drive. This means that the Floquet steady state here is in fact a quasi-steady state. In contrast, as shown below, after spatial dimensions are involved, temporal patterns are more efficiently stabilized to a true steady state. We deem this as a key difference between the well-known 0D single parametric oscillator and the more nontrivial multidimensional space-time crystal. (See Methods for a quantitative comparison between the 0D and 1D cases.)

\noindent\textbf{Effectively 1D single-layer long junction.} The investigation above merely recovers the Floquet dynamics of a nonlinear parametric oscillator. However, by extending the studies into 1D, we find that even if the modulation still follows Eq.~(\ref{eqn:jc_mod}), the system permits spontaneous translational symmetry breaking. The system tends to pick favorable modes within a broadened modulation band that can be higher than $2\fpl$. Figure~\ref{fig:1D} shows our calculations for an effectively 1D single-layer long Josephson junction in BSCCO with $L=25~\mu$m and $W=2$~$\mu$m. The spontaneously developed space-time crystalline order, especially, with rhombohedral unit cells, is remarkable. This symmetry is classified as $C_2m_xm_t$ in Ref.~\cite{Xu18}. To our knowledge, BSCCO provides the first material realization of this new symmetry class. Fig.~\ref{fig:1D}a gives the space-time trace $j_z(x,t)$ at $\fmod=140$~GHz and $\Amod=0.1$. Fig.~\ref{fig:1D}c gives its spatiotemporal Fourier transform in terms of the amplitude $|j_z(\beta,f)|$, where $\beta$ is the wavenumber (reciprocal of wavelength) along $x$. It shows a strong peak at the half-harmonic frequency $\fmod/2$ and a wavenumber corresponding to mode index $l=9$. This is manifested by the line projection of the Fourier amplitude across the half-harmonic peak versus frequency in Fig.~\ref{fig:1D}b and versus wavenumber in Fig.~\ref{fig:1D}d, respectively. We then define a spatiotemporal correlation function,
\begin{equation}
g(\delta x,\delta t) = \frac{\langle j_z(x,t) j_z(x+\delta x, t+\delta t)\rangle_{x,t}}{\sqrt{\langle j_z^2 (x,t)\rangle_{x,t}\langle j_z^2 (x+\delta x,t+\delta t)\rangle_{x,t}}},
\end{equation}
where the average is taken for both space and time. Fig.~\ref{fig:1D}e gives a color plot for $g(\delta x,\delta t)$, displaying nearly perfect sinusoidal oscillations along both $\delta x$ and $\delta t$. To see the dependence of space-time crystalline order with the modulation parameters, we define a coupled spatiotemporal order parameter,
\begin{equation}
\begin{split}
& \Delta_{xt} = \frac{1}{2} \left| \frac{\langle j_z(0,t) j_z(L, t+2\nu \Tmod)\rangle_{t,\nu}}{\sqrt{\langle j_z^2 (0,t)\rangle_{t,\nu}\langle j_z^2 (L,t+2\nu \Tmod)\rangle_{t,\nu}}} \right.\\
& \left. - \frac{\langle j_z(0,t) j_z(L, t+(2\nu+1) \Tmod)\rangle_{t,\nu}}{\sqrt{\langle j_z^2 (0,t)\rangle_{t,\nu}\langle j_z^2 (L,t+(2\nu+1) \Tmod)\rangle_{t,\nu}}} \right| ,
\end{split}
\end{equation}
as a natural extension of purely the temporal order parameter Eq.~\ref{eq:OrderT}. This order parameter cannot be separated into a direct product of a spatial order parameter and temporal order parameter. It correlates the out-of-plane supercurrents at the most far distant points in space -- the boundaries at $x = 0$ and $x = L$ in 1D (and diagonally opposite corners in higher dimensions), consistent with our rectangular spatial system geometry. We find, because of the particular complexity of our system, this definition of order parameter (or quasi order parameter) is practically the most rational and universal choice, owing to the fact that whenever a spatiotemporal order emerges and collective modes have formed, the long-range correlation function across the system must be large.

In Fig.~\ref{fig:1D}f, we plot the phase diagram of $\Delta_{xt}$ versus $\fmod$ and $\Amod$. Fig.~\ref{fig:1D}g gives a linescan of Fig.~\ref{fig:1D}f when $\Amod=0.1$. We can see that the half-harmonic response now extends into a broad frequency band from 120 to 160~GHz. Comparing with the 0D case, exhibiting an only 10~GHz wide band near and slightly below $2\fpl$, the 1D system is more robust to the choice of driving source. The peaks with $\Delta_{xt} > 0.5$ seen in Fig.~\ref{fig:1D}g correspond to mode indices $l$ between 3 and 10. Note that the fundamental mode with $l = 0$ when $\fmod\approx 2\fpl$, which one may anticipate to show up, is suppressed.

\begin{figure*}[htb]
\centerline{\includegraphics[scale=0.5]{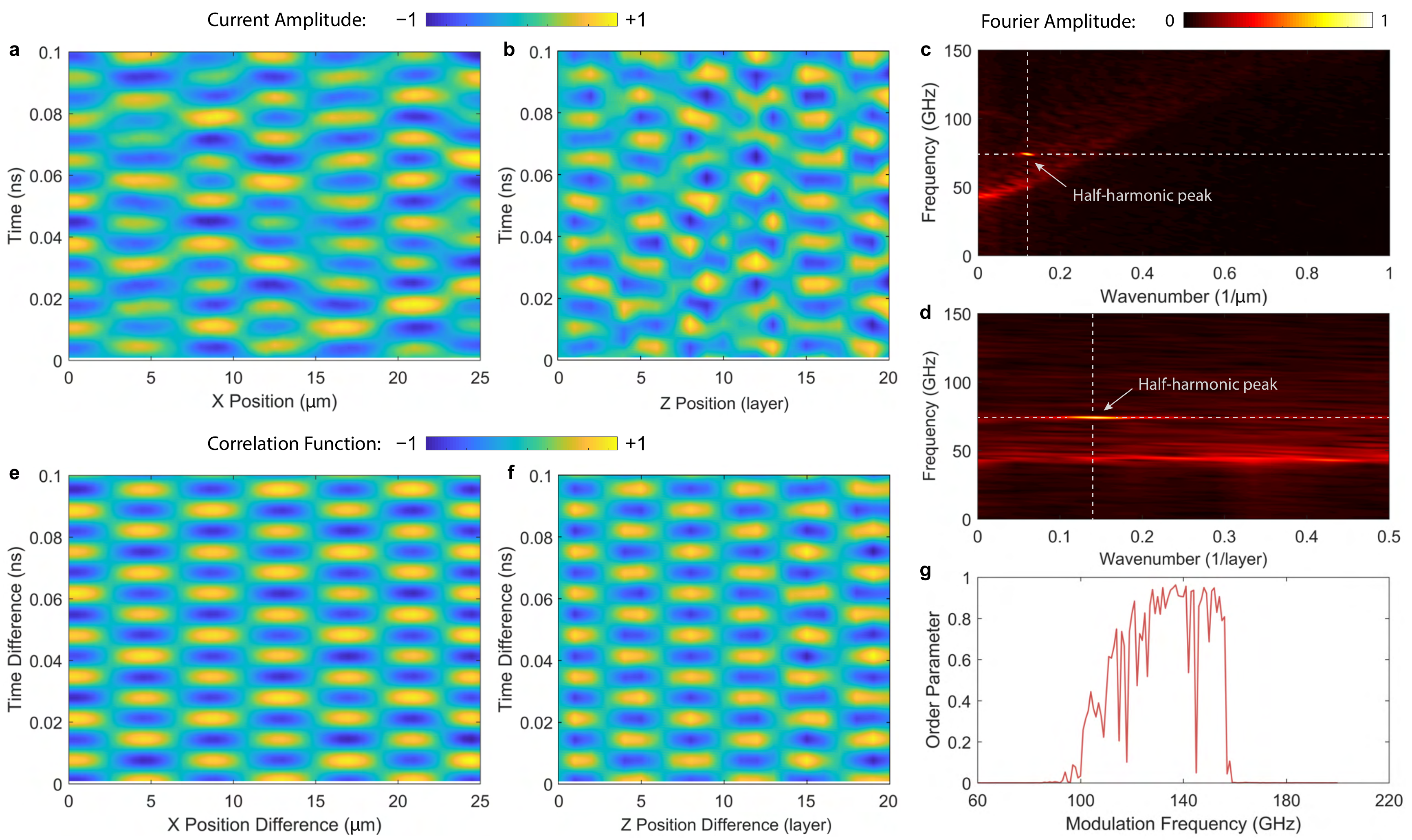}}
\caption{\textbf{Calculation on an effectively 2D multi-layer long Josephson junction stack in BSCCO at 4~K under parametric modulation.} The junction length is $L=25$~$\mu$m, width is $W=2$~$\mu$m, and layer number is $N=20$. \textbf{a} Space-time trace of $j_z(x,0,t)$ in the $xt$ plane when $z=0$ in the Floquet steady state under the modulation frequency $\fmod=148$~GHz and modulation amplitude $\Amod=0.1$. \textbf{b} Space-time trace of $j_z(0,z,t)$ in the $xt$ plane when $x=0$. \textbf{c} Fourier transform to $j_z(x,0,t)$ plotted in the magnitude $|j_z(\beta,0,f)|$, where $\beta$ is the wavenumber along $x$. \textbf{d} Fourier transform to $j_z(0,z,t)$ plotted in the magnitude $|j_z(0,\gamma,f)|$, where $\gamma$ is the wavenumber along $z$. \textbf{e} Space-time-averaged correlation function $g(\delta x,\delta z,\delta t)$ in the $\delta z=0$ plane. \textbf{f} Space-time-averaged correlation function in the $\delta x=0$ plane. The corresponding mode indices are $l=6$ and $q=6$. \textbf{g} Spatiotemporal order parameter $\Delta_{xzt}$ versus the modulation frequency $\fmod$ when $\Amod=0.1$ (with fixed $\fpl$, temperature $T=$~4K, and thermal noise). It shows a more continuously filled parametric excitation band (compared with the 1D case) spanning 100 to 160~GHz for $\Amod=0.1$. For $\Amod=0.1$ the correction factor $C=0.91$.}
\label{fig:2D}
\end{figure*}

\noindent\textbf{Effectively 2D multi-layer long junction stack.} Next, we include the spatial dimension along $z$ by increasing the junction number $N$ from 1 to 20. This dimension is distinct from $x$ and $y$, and is unique for the naturally grown BSCCO crystal compared with the traditionally fabricated planar Nb or Al junction arrays \cite{Welp13,Hu10}. Figure~\ref{fig:2D} gives our calculated results in this scenario. Fig.~\ref{fig:2D}a and \ref{fig:2D}b show the space-time traces in the $xt$ and $zt$ plane at the fixed points $z=0$ and $x=0$, respectively. The modulation frequency is $\fmod=150$~GHz and the amplitude is $\Amod=0.1$. 2D crystalline order emerges in both directions. In Fig.~\ref{fig:2D}c and \ref{fig:2D}d, we show the space-time Fourier transform for \ref{fig:2D}a and \ref{fig:2D}b respectively and plot the Fourier amplitudes in the wavenumber-frequency planes. In addition to the half-harmonic generation in $t$, the system spontaneously chooses the mode $l=6$ in $x$ and $q=6$ in $z$. We can define the further generalized spatiotemporal correlation function,
\begin{equation}
\begin{split}
& g(\delta x,\delta z,\delta t) = \\
& \frac{\langle j_z(x,z,t) j_z(x+\delta x, z+\delta z, t+\delta t) \rangle_{x,z,t}}{\sqrt{\langle j_z^2(x, z, t) \rangle_{x,z,t}\langle j_z^2(x+\delta x, z+\delta z, t+\delta t) \rangle_{x,z,t}}},
\end{split}
\end{equation}
by taking averages in all dimensions and noting $z=sn$ and $\delta z = s\delta n$. In Fig.~\ref{fig:2D}e and \ref{fig:2D}f, we plot $g(\delta x, 0,\delta t)$ and $g(0, \delta z, \delta t)$, which again verifies the nearly perfect crystalline order between each pair of space-time dimensions. We can also define a further generalized coupled spatiotemporal order parameter by
\begin{equation}
\begin{split}
& \Delta_{xzt} = \frac{1}{2} \left| \frac{\langle j_z(0,0,t) j_z(L, H, t+2\nu \Tmod)\rangle_{t,\nu}}{\sqrt{\langle j_z^2 (0,0,t)\rangle_{t,\nu}\langle j_z^2 (L,H,t+2\nu \Tmod)\rangle_{t,\nu}}} \right.\\
& \left. - \frac{\langle j_z(0,0,t) j_z(L, H, t+(2\nu+1) \Tmod)\rangle_{t,\nu}}{\sqrt{\langle j_z^2 (0,0,t)\rangle_{t,\nu}\langle j_z^2 (L,H,t+(2\nu+1) \Tmod)\rangle_{t,\nu}}} \right| ,
\end{split}
\end{equation}
and plot its dependence on $\fmod$ at the fixed $\Amod=0.1$, as shown in Fig.~\ref{fig:2D}g. With the addition of the $z$-dimension, the system shows an almost continuous modulation bandwidth between 100 and 160~GHz.

\noindent\textbf{3D multi-layer long and wide junction stack.} Finally, we study the full 3D situation by assigning the $y$ dimension a finite width $W=5$~$\mu$m, and allowing for spatial variations of the Josephson phase differences along $y$. Such a width can be easily obtained in experiments \cite{Hu10}. Since in BSCCO, $x$ and $y$ are equivalent directions, we do not expect drastic changes of physics from the 2D case except for further enhanced robustness of space-time crystalline order due to the increased dimensionality. Our calculation does verify this expectation. Instead of repeating same sets of plots as above, we consider a possible experimental setup (see Methods) and perform the 3D calculation with experimental parameters.

\begin{figure}[htb]
\centerline{\includegraphics[scale=0.5]{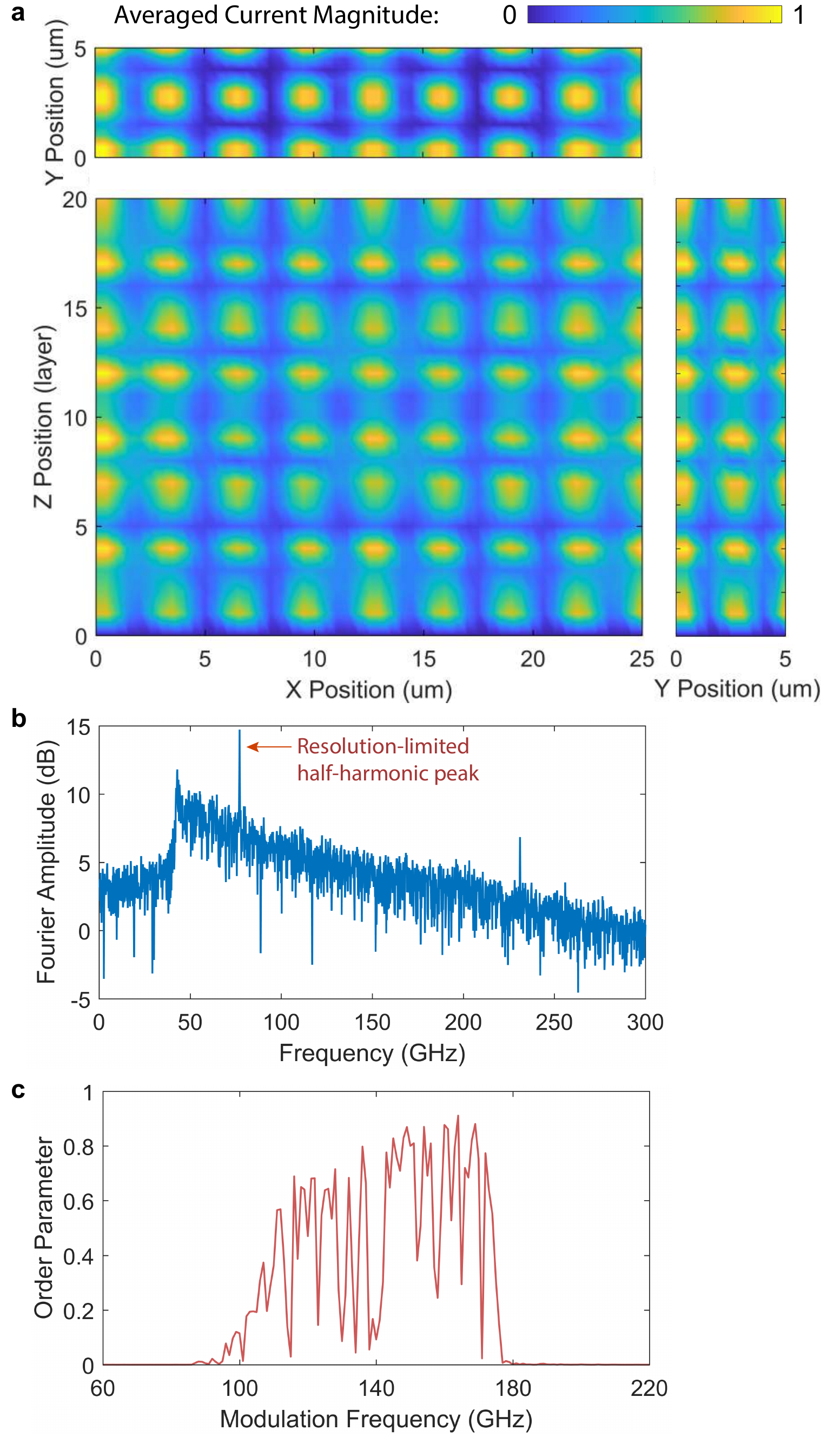}}
\caption{\textbf{Calculation on a 3D multi-layer long and wide Josephson junction stack in BSCCO under periodic laser illumination.} The junction length is $L=25$~$\mu$m, width is $W=5$~$\mu$m, and layer number is $N=20$. \textbf{a} An example of expected spatial profiles of $\langle j_z^2(x,y,z,t)\rangle/\jc^2$ in the Floquet steady state after being averaged over $(z,t)$, $(x,t)$ and $(y,t)$ and plotted in the $xy$, $yz$ and $xz$ planes, respectively. The modulation frequency is $\fmod=154$~GHz, the laser power is $P=0.5$~mW, and the bath temperature is $\TB=4$~K. \textbf{b} Expected Fourier spectrum of $j_z(x,y,z,t)$ on the edge of the sample (0,0,0) where it can convert to electromagnetic radiation and be detected. The spectrum shows the highest spark exactly at the half-harmonic frequency 77~GHz. \textbf{c} Expected parametric excitation band for the spatiotemporal order parameter for various modulation frequencies $\fmod$ at the fixed laser power 0.5~mW. This spectrum is further broadened (compared with the 2D case).}
\label{fig:3D}
\end{figure}

We assume a laser beam, modulated in intensity at a frequency $f_{\rm mod}$, is absorbed by the BSCCO stack where it deposits an ac power $P_{\rm ac} = P_{\rm ac,0} \sin^2(2\pi f_{\rm mod}t/2)$. If the BSCCO crystal is considered at thermal equilibrium, the result of the laser irradiation would just be a rise in temperature, without appreciable ac oscillations in the stack temperature, or phononic temperature. However, if the crystal is cooled sufficiently, one may see oscillations in the electronic temperature \cite{Sobolewski98}, which is what we need.

In our calculation, we assume that the stack is glued to a 30\,$\mu$m thick substrate, whose lower surface is kept at the bath temperature $\TB$, which in practice is often lower than the temperature $T$ of the junctions. The thickness of the glue layer is 30\,$\mu$m. We assume a local equilibrium between the electrons and phonons in the crystal. To mimick oscillations in the electronic temperature we use extremely low values of the heat capacities of 5\,J/m$^3$K. We find, when starting at a bath temperature of 4\,K and using $P_{\rm ac,0}$ = 0.5\,mW, a time-averaged stack temperature of 29\,K and an oscillation amplitude of 5.3\,K, leading to a 10$\%$ modulation of $j_{\rm c}$, a 2$\%$ modulation of $n_{\rm s}$, a 20$\%$ modulation of $\rho_c$ and a 10$\%$ modulation of $\rho_{ab}$. The time-averaged stack temperature is actually not constant across the stack but decays from 29.2\,K in the center of the stack to 26.8\,K at its edges. Fig.~\ref{fig:3D}a displays projections of the time-averaged intensity  $g(0,0,0,0) = \langle j_z^2(x,y,z,t) \rangle/\jc^2$ of the normalized out-of-plane current densities $j_z$ flowing in the stack. The projections are to the $(x,y)$, $(x,z)$ and $(y,z)$ planes, respectively.  The modulation frequency is $f_{\rm mod}$ = 154\,GHz. The mode indices of the standing wave pattern are $(l,m,q)$ = (8,2,13) and the oscillation frequency of this mode is 77\,GHz.

Experimentally, to visualize the spatial order in the $xy$ plane, one can use either near-field measurements of the ac electric fields near the stack surface \cite{Stewing08,Adam11} or low-temperature scanning laser microscopy \cite{Wang09a}. The latter technique has already been used to visualize standing electromagnetic waves in BSCCO stacks used for generation of THz radiation. In these experiments, the sample is usually current biased and a laser beam focused on the sample induces local heating which in turn changes the local electric properties. As a result, the global voltage across the sample changes. These global voltage changes, which are related to the local time average $\langle j_z^2(x,y,t) \rangle$ in our case, can be mapped by scanning. The typical resolution is around 1~$\mu$m, allowing to resolve many of the standing wave patterns along $x$ and $y$. Fourier-transform infrared spectroscopy (FTIR) can be used to verify the predicted spatiotemporal ordered states and resolve the half-harmonic generation of the modulation frequency $\fmod$ in time. Fig.~\ref{fig:3D}b shows the expected spectrum. Fig.~\ref{fig:3D}c shows the calculated parametric modulation band as an indicator of the driving robustness.

\noindent\textbf{Discussion}

One may quest whether the spatiotemporal order in BSCCO can be extended to spatially arbitrarily large stacks. With increasing stack size, particularly along the $z$ direction, the number of resonant modes drastically increases, with a large number of modes that are nearly degenerate. A strong mode competition can occur in this case, eventually leading to a decreased mode stability and/or to chaotic behavior. Our preliminary simulations for larger junction numbers indeed indicate this trend. Therefore, a junction number of order 20 may already be close to the maximum in order not to destroy the space-time order. More in-depth studies are needed along this direction.

Besides, we have not specifically mentioned the effect of bath temperature, which is proportional to the thermal fluctuation that may cause mode unlocking. At low temperatures around 4~K, within the timescales of our simulations we have not observed any mode-unlocking instability. However, at high temperatures, \eg, above 30~K, we have indeed found that increased thermal fluctuations cause mode unlocking.

A fundamentally intriguing question is how the system will behave in the true thermodynamic limit, \ie, when the degrees of freedom, volume, and time all go to infinity. According to our simulations for a finite system, when the system degrees of freedom increase, the long-wavelength soft modes tend to be suppressed, likely due to the random destruction interference under strong drive. We thus speculate that in the true thermodynamic limit, only the modes with a spatial period on the order of the Josephson length $\lj$ may survive.

In summary, we have shown theoretically that the high $T_\text{c}$ superconductor BSCCO is a natural candidate for a classical space-time crystal (STC), owing to its property to intrinsically form a stack of long (and wide) Josephson junctions. Under a temporally periodic and spatially uniform modulation of the Josephson critical current density, BSCCO can display space-time crystalline order of the supercurrents that breaks the continuous translational symmetry in both space and time. If the size of BSCCO is comparable or smaller than the Josephson length in all directions, the system has effectively zero dimension and merely exhibits the well-known parametric oscillations in time and demands an infinite time to evolve towards a (quasi-)Floquet steady state. In contrast, finite (one to three) spatial dimensions ensure a rapid convergence of the system towards the real STC state.
The modulation frequency and amplitude span a nonequilibrium phase diagram for a so-defined spatiotemporal order parameter. The phase diagram shows clear boundaries, outside of which the system is either disordered or chaotic. With increasing spatial dimensions, the system shows increasing stability manifested by a broadened modulation bandwidth.

Our calculations are all based on actual properties of BSCCO and so can be used to predict experiments. We have envisioned an experimental scheme to realize the BSCCO STC by using a near-infrared laser with a repetition rate of 100~--~200~GHz to modulate the Josephson current density. The signature of STC can be indirectly confirmed by measuring the half-harmonic emission spectrum via FTIR or directly visualized by the well-developed low-temperature scanning laser microscopy. If experimentally evidenced, this new kind of condensed-matter space-time crystals not only enriches the classical and quantum many-body physics in nonequilibrium, but also holds promise for unprecedented applications in, \eg, tunable emitters and parametric amplifiers at terahertz frequencies.

\noindent\textbf{Methods}

\noindent\textbf{Calculation scheme.} The basic model describing the electromagnetic and thermal properties of IJJ stacks is given in Ref.~\cite{Rudau15,Rudau16}. Here we give a short summary, with an emphasis on the geometry used for the present paper.

We consider a rectangular stack of $N$ IJJs, each having a thickness (layer period) $s$ = 1.5\,nm. The thickness of the superconducting layers (CuO$_2$ double layers) is $d_{\text{s}}$ = 0.3\,nm and the thickness of the barrier layers is $d_{\text{i}}$ = 1.2\,nm. Like in the sketch of Fig.~\ref{fig:schematic} the stack of length $L$ and width $W$ shall be a stand-alone stack, \ie, without a BSCCO base crystal underneath.  The critical temperature of the stack is $T_{\rm c}$ = 85\,K.
The electromagnetic part of the circuit is formulated in terms  of in-plane and out-of-plane current densities flowing, respectively, along the $n$th CuO$_2$ layer or across the $n$th junction in the stack.

The out-of-plane current densities $j_{{z},n}$ across the $n$th IJJ between the CuO$_2$ planes $n$ and $n-1$ consist of Josephson currents with critical current density $j_{{\rm c}}$, (ohmic) quasiparticle currents with resistivity $\rho_{c}$ and displacement currents with dielectric constant $\varepsilon$. We also add Nyquist noise created by the quasiparticle currents. The in-plane currents along the $n$th CuO$_2$ plane consist of a superconducting part, characterized by a Cooper pair density $n_{{\rm s}}$ and a quasiparticle component with resistivity $\rho_{ab}$. The model parameters depend on temperature, as plotted in detail in Ref. \onlinecite{Rudau15}. Then, the electric part of the circuit is described by sine-Gordon-like equations for the Josephson phase differences $\gamma_{ n}(x,y)$ in the $n$th junction of the IJJ stack:
\begin{equation}
\label{eq:sigo_segment}
\begin{split}
sd_{\rm s}\nabla\left(\frac{\nabla\dot{\gamma}_n}{\rho_{ab}}\right)
%+d_{\rm s}\nabla(j^{\rm N}_{{x},l+1}-j^{\rm N}_{{x},l})
+ \lambda_{k}^2 \nabla(n_\text{s}\nabla\gamma_n) = \\
\left(2+\frac{\lambda_{k}^2n_s}{\lambda_{\rm c}^2}\right)j_{z,n}-j_{z,n+1}-j_{z,n-1}.
\end{split}
\end{equation}
Here, $n$ = $1\dots N$, $\nabla = (\partial/\partial x, \partial/\partial y)$ and
$\lambda_{k} = [\Phi_0 d_{\rm s}/(2\pi\mu_0j_{\rm c0}\lambda_{ab0}^2)]^{1/2}$, with the in-plane London penetration depth $\lambda_{ab0}$ at 4~K and the magnetic permeability $\mu_0$. $\lambda_{\rm c} = [\Phi_0/(2\pi\mu_0j_{\rm c0}s)]^{1/2}$  is the out-of-plane penetration depth. $j_{\rm c0}$ is the 4~K Josephson critical current density and $\rho_{c0}$ is the BSCCO $c$ axis (subgap-)resistivity at 4~K. $\Phi_0$ is the flux quantum. For the out-of-plane current densities,
\begin{equation}
\label{eq:RCSJ}
j_{{z},n} = \beta_{\rm c0} \ddot{\gamma}_n + \frac{\dot{\gamma}_n}{\rho_{{c},n}} + j_{\rm c} \sin(\gamma_n) + \eta_{{z},n},
\end{equation}
with $\beta_{\rm c0} = 2\pi j_{\rm c0}\rho_{c0}^2\varepsilon\varepsilon_0s/\Phi_0 $; $\varepsilon_0$ is the vacuum permittivity and $\eta_{{z},n}$ are the out-of-plane noise current densities. Time is normalized to $\Phi_0/2\pi j_{\rm c0}\rho_{c0}s$, resistivities to $\rho_{c0}$ and current densities to $j_{\rm c0}$.

For the electrical parameters we use the 4~K values $\rho_{c0} $ = 1100~$\Omega$~cm, $j_{\rm c0} = 250$\,A/cm$^2$, $\lambda_{\rm ab0}$ = 260\,nm, and further $\rho_{ab} (T_{\rm c})$ = 100\,$\mu \Omega$cm and $\varepsilon = 13$, yielding $\lambda_c$ =  264\,$\mu$m, $\lambda_k$ = 1.52\,$\mu$m and $\beta_{\rm c0}$ = 1.5 $\times$ 10$^5$.
We will further make use of the Josephson length $\lj$, obtained via $\lj^{-1} = 2\lambda_k^{-1}+\lambda_c^{-1}$. For our parameters $\lj \sim$ 1.1\,$\mu$m. For the temperature dependence of $j_{\rm c}$ we use a parabolic profile, $j_{\rm c} \propto 1-(T/T_{\rm c})^2$; for the temperature dependence of $\rho_{\rm c}$ see Ref. \onlinecite{Rudau15}.

Equations \ref{eq:sigo_segment} and \ref{eq:RCSJ} are discretized in space, with $X$ = 50 points in $x$ direction and $Y$ points in $y$ direction. For 3D simulations $Y = 10$ and for 2D simulations $Y = 1$. The equations are propagated in time using a 5th order Runge-Kutta method. For each pixel the normalized spectral density of $\eta_{{z},n}$ is $4\Gamma XY\rho_{c0}/\rho_{c}$, where $\Gamma = 2\pi k_{\rm B}T/I_{\rm c0}\Phi_0 $ is the noise parameter.

In the 0D-2D simulations discussed in the main paper, we consider a minimal model where we assume that the stack is at a fixed temperature $T$, with a homogeneous temperature distribution inside the stack. For these simulations we assume that the Josephson critical current density oscillates in time with an amplitude $\Amod$, $j_{\rm c}(t) =  j_{\rm c} [1+\Amod\cos(2\pi f_{\rm mod}t)]/(1+\Amod)$ and $j_{\rm c}(t)$ is homogeneous in space.

In the 3D simulations discussed in the main paper, we assume that some power $P_{\rm ac} = P_{\rm ac0}\cos^2(2\pi f_{\rm mod}t/2)$ is deposited in the stack and solve Eqs. \ref{eq:sigo_segment} and \ref{eq:RCSJ} in combination with the heat diffusion equation $c{\rm d}T/{\rm d}t=\nabla(\kappa\nabla T)+q_s$, with the specific heat capacity $c$ and the (anisotropic and layer dependent) thermal conductivity $\kappa$. The power density $q_s$  for heat generation in the stack results from Joule heating due to the electrical part of the circuit and from the ac power deposited by the laser.  For the simulations we assume that the stack is glued to a 30\,$\mu$m thick substrate with lateral dimensions $2L\times2W$, whose lower surface is kept at the bath temperature $\TB$. The thickness of the glue layer is 30\,$\mu$m. To mimick oscillations in the electronic temperature we use extremely low values for the heat capacities of 5\,J/m$^3$~K. The thermal equations are discretized in space, using $2X$ points along $x$ and $2Y$ points along $y$ and are propagated in time using a 5th order Runge-Kutta method.

\begin{figure}[htb]
\centerline{\includegraphics[scale=1]{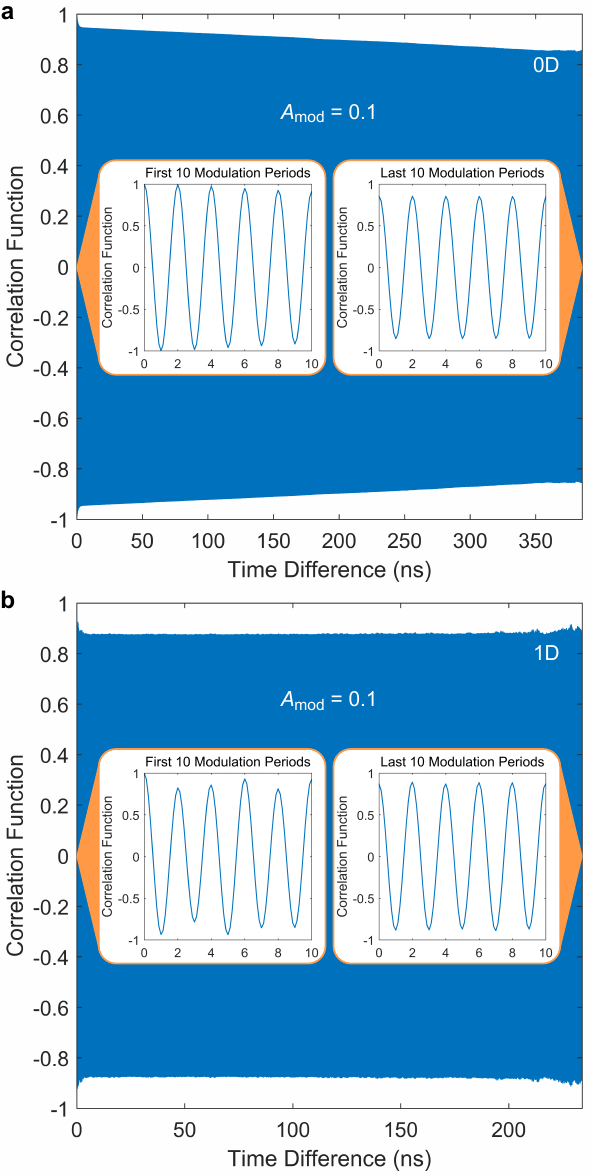}}
\caption{\textbf{Comparison between the calculated temporal correlation function for 0D and 1D Josephson junctions in BSCCO.} The time difference $\delta t$ takes up to 32768 cycles of modulation period $\Tmod=1/\fmod$ and the modulation amplitude is $\Amod=0.1$. \textbf{a} 0D case with $\fmod=85$~GHz. The insets show slightly decayed oscillation amplitudes for the first and last 10 modulation periods. \textbf{b} 1D case (evaluated at a specific point with the space difference $\delta x=0$) with $\fmod=140$~GHz. The insets show nearly the same oscillation amplitudes for the first and last 10 modulation periods.}
\label{fig:0D1DCompare}
\end{figure}

It is important to note the difference between 0D and finite spatial dimensions. For 0D, within our calculated time scale, the system has never reached a true steady state, \ie, a desired time-crystal phase. As shown in Fig.~\ref{fig:0D1DCompare}(a), after the system passes the transient state and enters the Floquet (quasi-)steady state, the temporal correlation function persistently decays even by a time difference of 32768 cycles of modulation period $\Tmod=1/\fmod$, with $\fmod=85$~GHz. In contrast, for 1D, as shown in Fig.~\ref{fig:0D1DCompare}(b), the temporal correlation (evaluated at a specific point with zero space difference) establishes quickly and remains the same up to 32768 cycles of $\Tmod=1/\fmod$, with $\fmod=140$~GHz.

\begin{figure}[htb]
\centerline{\includegraphics[scale=0.45]{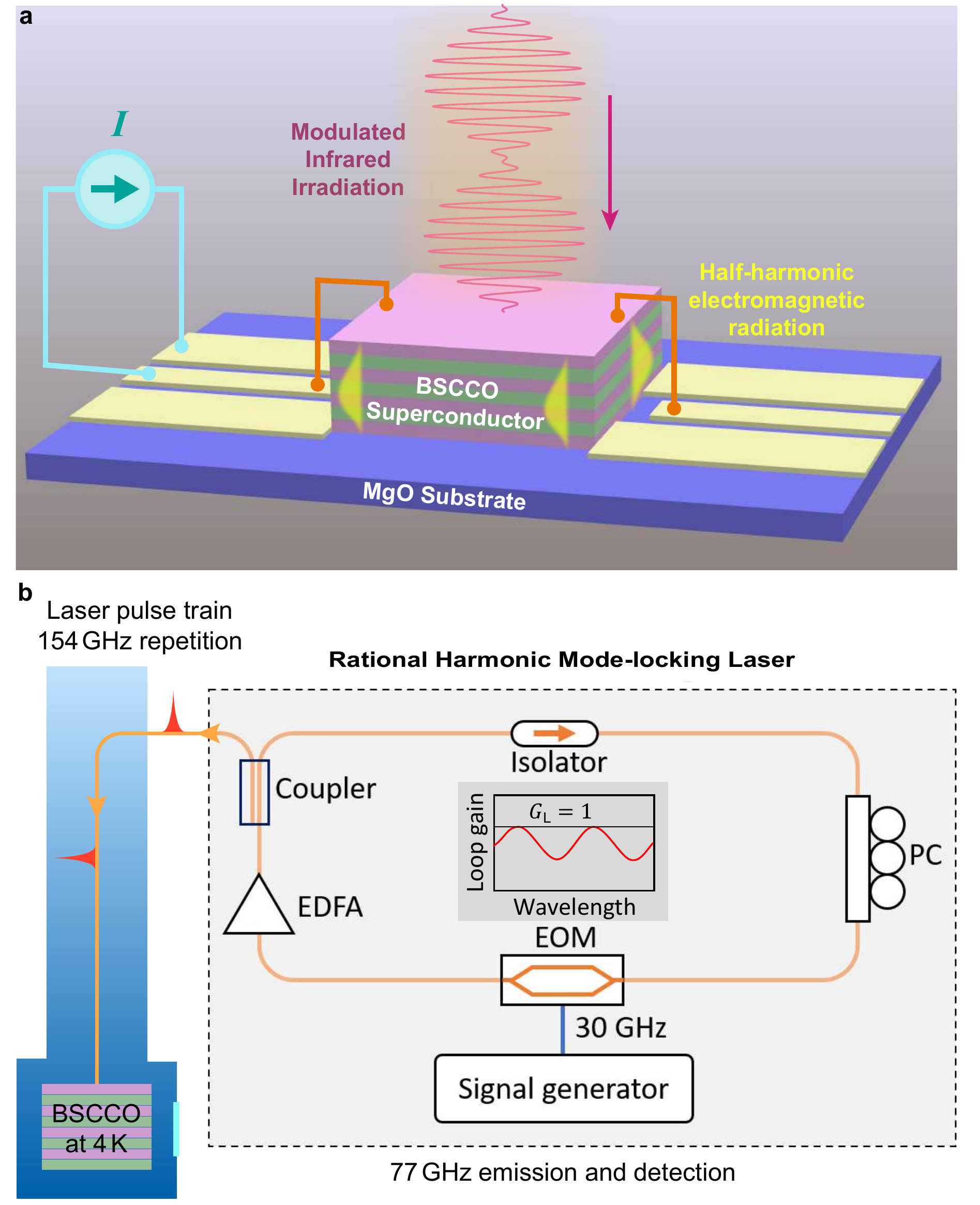}}
\caption{\textbf{Experimental scheme of a parametrically modulated stack of multi-layer long and wide Josephson junctions in BSCCO.} \textbf{a} Schematics of a laser-pumping induced space-time crystal in BSCCO. \textbf{b} Proposed setup to achieve 154~GHz laser modulation from 30.8~GHz using the rational harmonic generation technique and to detect the half-harmonic generation either with a Fourier-transform infrared spectroscopy or autocorrelation setup. Inset shows the loop gain profile with a sinusoidal line shape originating from the EOM modulation. The amplitude condition for lasing is satisfied when the gain profile equals unity ($G_\text{L}=1$).}
\label{fig:exp}
\end{figure}

\noindent\textbf{Experimental proposal.} A schematic experimental design to generate periodic modulations of the junction parameters is given in Fig.~\ref{fig:exp}a. The major challenge is to achieve a very high repetition rate $100$ to $200$~GHz and a strong power so as to induce about 10\% change of $\jc$, \ie, $\Amod=0.1$. One may consider raising the bath temperature $\TB$ close to $\tc$ to reduce $\fpl$ and thus alleviate the frequency requirement. However, the simultaneously increased thermal noise significantly suppresses the formation of space-time crystalline order. In fact, in our simulations we could not find a substantial spatiotemporal order above $\sim$40\,K even in the 3D case.

Laser pulses with a high repetition rate can be achieved by adopting the rational harmonic mode-locking (RHML) laser technique \cite{yoshida199680,das1997rational}. As shown in Fig.~\ref{fig:exp}b, the construction of a RHML laser consists of a closed-loop fiber ring as the optical cavity. An erbium doped-fiber amplifier (EDFA) is plugged inside the fiber ring to provide gain for lasing in the telecommunication C-band. In general, the lasing frequency (wavelength) of a ring laser is determined by two conditions: (1) amplitude condition: loop gain $G_\text{L}=1$; (2) phase condition: loop phase $\varphi_\text{L}=2\pi M$ with $M$ being an arbitrary integer. The amplitude condition is determined by a lithium niobate electro-optical modulator (EOM) inserted into the ring, which modulates the loop gain profile and consequently determines the lasing frequency, while the phase condition is controlled by carefully selecting the modulation frequency. When the EOM operates at a frequency $f_\text{m}=\left(s+1/p\right)f_\text{c}$, where $s$ and $p$ are integers, and $f_\text{c}$ is the fundamental frequency of the fiber ring cavity which is determined by the loop length of the fiber ring, the repetition rate of the laser pulses can be obtained as $f_\text{r}=(sp+1)f_\text{c}$. As an estimation, a 5~m long single mode (SM) fiber loop gives $f_\text{c}=40$~MHz. Choosing $s=770$ and $p=5$, a high repetition rate of $f_\text{r}=154$~GHz can be obtained, while the required modulation frequency $f_\text{m}=30.8$~GHz is readily accessible using standard EOM and microwave sources. In practice, the polarization of the laser is determined by the polarization-maintaining elements (such as the EOM) inside the ring. The polarization controller (PC) is used to optimize the polarization condition for these elements, and an optical isolator is used to reinforce unidirectional propagation of the laser light inside the ring. Most importantly, such a configuration ensures that the modulation signal, laser repetition rate, and half-harmonic generation are all at different frequencies, and therefore potential crosstalk in the characterization of the generated half-harmonic signal can be drastically reduced.

\noindent\textbf{Data Availability}

The authors declare that all data supporting the findings of this study are available within the paper.

\noindent\textbf{Code availability}

Computer codes are available from the corresponding authors upon reasonable request.

\noindent\textbf{Acknowledgements}

This work was performed in part at the Center for Nanoscale Materials, a U.S. Department of Energy Office of Science User Facility, and supported by the U.S. Department of Energy, Office of Science, under Contract No. DE-AC02-06CH11357. E. D., D. K., and R. K. acknowledge support by the Deutsche Forschungsgemeinschaft via project KL930-13/2, and the EU-FP6-COST Action NANOCOHYBRI (CA16218).

\noindent\textbf{Author Contributions}

R. K. and D. J. conceived the idea and led the project. R. K. performed the numerical calculation supported by E. D.
X. Zhou analyzed the results and developed the experimental design. X. Zhang and D. K. provided additional advices and joined discussions. All authors contributed to the manuscript writing. R. K. and X. Zhou contributed equally to this work.

\noindent\textbf{Competing interests}

Authors declare no competing interests.

\bibliography{IJJ_Xtal}

\end{document}